\documentclass[conference]{IEEEtran}
\IEEEoverridecommandlockouts
\usepackage{cite}
\usepackage{amsmath,amssymb,amsfonts}
\usepackage{algorithmic}
\usepackage{graphicx}
\usepackage{textcomp}
\usepackage{cleveref}

\usepackage{xcolor}
\usepackage{amsmath,amssymb}
\usepackage[flushleft]{threeparttable}
\usepackage{array,booktabs,makecell}
\usepackage[font=small]{caption}
\def\BibTeX{{\rm B\kern-.05em{\sc i\kern-.025em b}\kern-.08em
    T\kern-.1667em\lower.7ex\hbox{E}\kern-.125emX}}
\begin{document}

\title{OpenSpike: An OpenRAM SNN Accelerator\\
}

\author{\IEEEauthorblockN{Farhad Modaresi}                   
\IEEEauthorblockA{\textit{Dept. of Electrical Engineering} \\
\textit{Allameh Mohaddes Nouri University}\\
Nur, Mazandaran, Iran \\
f.modaresi@mohaddes.ac.ir}
\and
\IEEEauthorblockN{Matthew Guthaus}
\IEEEauthorblockA{\textit{Dept. of Computer Science and} \\
\textit{Engineering, UC Santa Cruz}\\
Santa Cruz, CA, United States \\
mrg@ucsc.edu}
\and
\IEEEauthorblockN{Jason K. Eshraghian}
\IEEEauthorblockA{\textit{Dept. of Electrical and Computer} \\
\textit{Engineering, UC Santa Cruz}\\
Santa Cruz, CA, United States \\
jeshragh@ucsc.edu}
}

\maketitle

\begin{abstract}
This paper presents a spiking neural network (SNN) accelerator made using fully open-source EDA tools, process design kit (PDK), and memory macros synthesized using OpenRAM. 
The chip is taped out in the 130~nm SkyWater process and integrates over 1~million synaptic weights, and offers a reprogrammable architecture.
It operates at a clock speed of 40~MHz, a supply of 1.8~V, uses a PicoRV32 core for control, and occupies an area of 33.3~mm$^2$. The throughput of the accelerator is 48,262~images per second with a wallclock time of 20.72~$\mu$s, at 56.8~GOPS/W.
The spiking neurons use hysteresis to provide an adaptive threshold (i.e., a Schmitt trigger) which can reduce state instability.
This results in high performing SNNs across a range of benchmarks that remain competitive with state-of-the-art, full precision SNNs. The design is open sourced and available online: \textit{https://github.com/sfmth/OpenSpike}

\end{abstract}


\section{Introduction}
The open source community has enabled advances in deep learning (DL) to proliferate over the past decade. Much of these advances have successfully been ported to biologically plausible spiking neural networks (SNNs). SNNs have been used to model brain function, and can perform DL using gradient-based or Hebbian learning rules, often on a CUDA backend \cite{knight2021larger, turner2022mlgenn, stimberg2018brian2genn, eshraghian2021training}. To achieve this, the training methods that have been popularized by DL (e.g., error backpropagation) have also been adopted to training SNNs to much success \cite{neftci2019surrogate, eshraghian2021training}.
The benefits of SNNs are limited when ported onto CUDA-accelerated backends as the underlying instruction-set is constrained to SIMD/SIMT (single-instruction multiple-data/thread) processing. 

Demonstrating the value of SNNs goes beyond measuring test set accuracy on toy problems. Evaluating the performance of SNNs extends to its efficiency, often in terms of energy, latency, or synaptic operations per second. Hardware such as Loihi and SpiNNaker offer power profiling tools to estimate the overhead of SNN workloads, and programming packages such as SpikingKeras can provide coarse estimates based on the number of operations in a model \cite{davies2018loihi, furber2014spinnaker, bekolay2014nengo}. These tools have lowered the barrier to access neuromorphic algorithm development and have provided promising empirical demonstrations of spike-based computing.

Neuromorphic research is still highly exploratory, and despite the various engineering benefits that have been demonstrated, 
the research community remain uncertain with what feature sets, neuron models, and neural computations should be integrated on-chip. On the one hand, why should we implement features in silicon that are not commonly used? On the other hand, how do we know what benefits such features have to offer without hardware baselines?

The recent expansion of open source electronic design automation (EDA) and VLSI tooling is primed to do for neuromorphic chip design what has already been done for DL \cite{edwards2020google, ghazy2020openlane}. The explosion of new DL methods as applied to SNNs only keeps coming \cite{bellec2020solution, henkes2022spiking, yang2022weak, stewart2022meta, eshraghian2022memristor, zhu2022tcja, perez2021sparse, venkatesha2021federated, elbtity2022aptpu}, and lowering the barrier to develop ASICs can ultimately profile these workloads in a way that can demonstrate or dispute energy and latency advantages \cite{zhou2022gradient, kang2021build, eshraghian2019analog}.

To guide the direction of neuromorphic research along an open and reproducible trajectory, this paper presents a fully open source SNN accelerator, including the process design kit (PDK), the tooling used to synthesize the design, and the memory macros used to store synaptic weights. Our design was done in the SkyWater~130nm (SKY130) process, cleared all pre-tape-out checks, and includes 1,059,840 synaptic weights, and operates at a clock speed of 40~MHz.
Small-scale SNNs (i.e., on the order of 1,000-10,000 neurons) tend to be memory-limited in their performance, so we have used OpenRAM macros in the design to promote further optimization of near-memory compute tooling.


\section{Neuron Model}
The most common neuron model used in SNNs trained via gradient descent is the leaky integrate-and-fire neuron model \cite{dayan2005theoretical, lapicque1907louis}. It offers a reasonable approximation of neurons while retaining simplicity. A discrete-time version can be represented by the following equation:

\begin{equation}
\label{eq:euler}
    u_t=\beta u_{t-1} + x_{t-1}w - z_{t-1}u_{\rm thr},
\end{equation}

\noindent where $u$ is the membrane potential of the neuron, $x$ is the input to the neuron, $w$ is the weight attached to $x$, $\beta$ is the decay rate of the membrane potential, and the subscript $t$ refers to time. When the membrane potential exceeds the threshold $u_{\rm thr}$, an output spike is generated:

\begin{equation} \label{eq:spike}
    z_t =
    \begin{cases}
      1,  & \text{\rm if $u_t > u_{\rm thr}$} \\ 
      0, & \text{otherwise.} \\
    \end{cases}  
\end{equation}

Introducing a Schmitt trigger to the thresholding action of the neuron has the following effect:
\begin{equation} \label{eq:spike}
    z_t =
    \begin{cases}
      1,  & \text{\rm if $u_t > u_{\rm thr}^h$ and $i_t = 0$} \\ 
      0, & \text{otherwise.} \\
    \end{cases}  
\end{equation}
\noindent where $i_t$ refers to spike inhibition:
\begin{equation} \label{eq:schmitt}
    i_t =
    \begin{cases}
      0,  & \text{\rm if $z_t = 0$ and $u_t < u_{\rm thr}^l$} \\
      1, & \text{otherwise.} \\
    \end{cases}  
\end{equation}

As shown in Refs. \cite{eshraghian2022memristor, eshraghian2022navigating, eshraghian2022fine}, SNNs are highly tolerant to weight quantization. In the extreme, the trainable weights of an SNN can be binarized to $w \in \{-1,+1\}$ with small performance degradation as performance is `propped up' by non-binarized variables, including membrane potential and time.

% ...

\section{Architecture of the Accelerator}
\begin{figure}[!t]
		\centering
		\includegraphics[width=\linewidth]{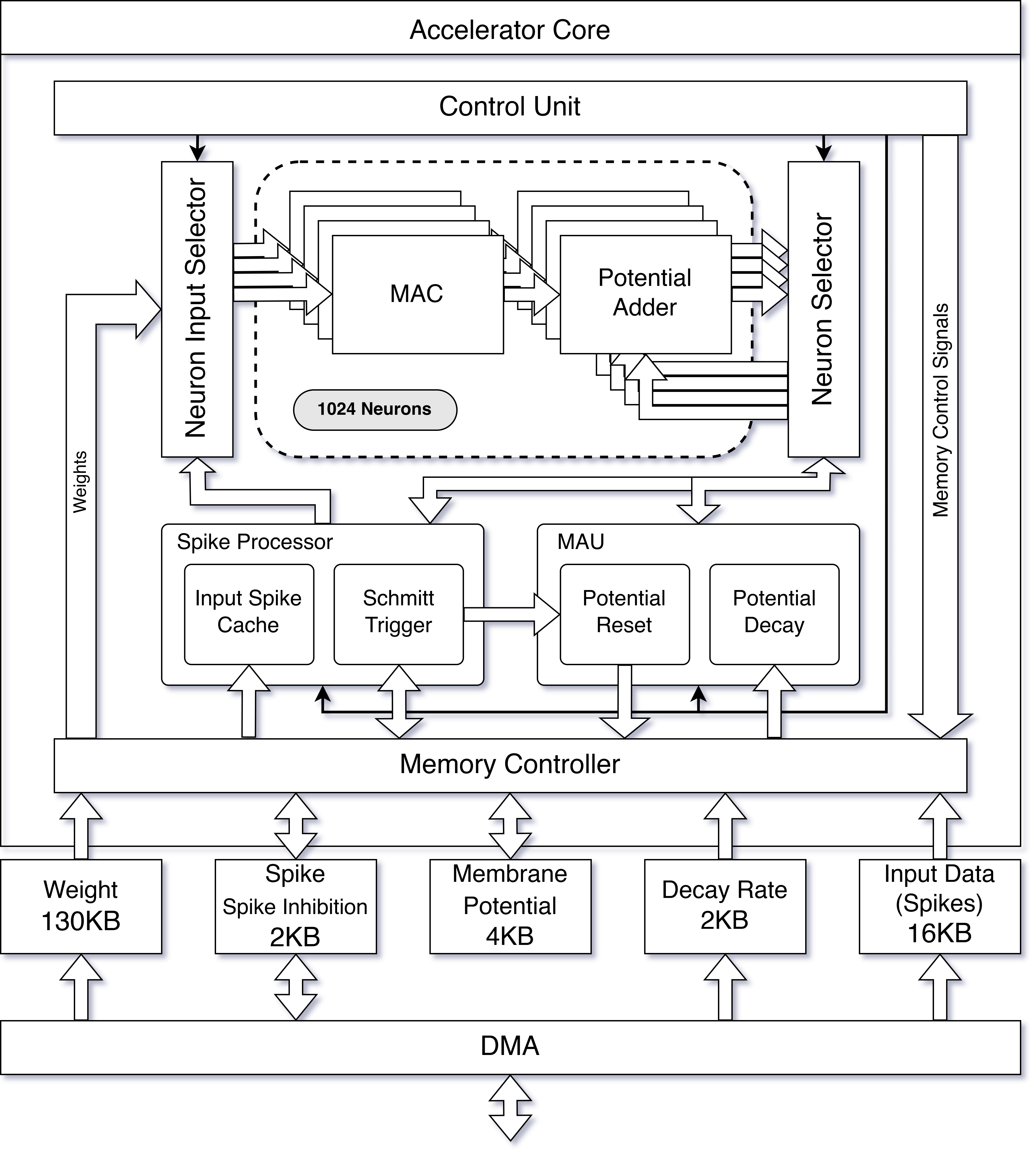}
		\caption{System architecture: Data flow between different modules of the OpenSpike core.}
		\label{fig:soc}
\end{figure}

We propose a time-multiplexed accelerator designed for the SKY130 process with an open-source PDK, synthesized and hardened using fully open-source EDA tools. The accelerator core is illustrated in \Cref{fig:soc}.
When carrying out neuromorphic computations, the memory complexity of DL and SNN workloads scales with $\mathcal{O}(n^2)$ where $n$ is the number of neurons, which is dominated by the network parameters. Fully-integrated accelerators on a monolithic substrate tend to be bottelenecked by weight access. 

To address this in our OpenSpike core, spiking neuron modules are re-used and time-multiplexed by parallelizing neuronal computations with the loading of weights, along with time-multiplexing inter-neuron computations. Neuron re-use optimizes for area with a marginal impact on latency. This is possible due to the parallelization of weight access and neuronal computations (i.e., state update, and spike triggering). Furthermore, using binarized weights balances the cost of weight reads from memory with neuron computations.


\subsection{Network Architecture}
To implement a neural network, our accelerator uses 1,024 hardware neurons to process the network one layer at a time. For example, in an SNN with an architecture of 1024-1024-10 dense connections across three layers, a total of 1,024 neurons would be needed on chip. A drawback to this approach is the need to save and load neuron data for each layer, but this can be performed in parallel with weight read-out with a marginal impact on performance.


\subsection{Neuron Processor}
Each of the 1,024 neurons consists of a multiply-accumulate (MAC) unit and a potential adder. The MAC unit sums the product of inputs $x$ with weights $w$ over time, and stores the result in a register. The final result is then passed to the potential adder which accumulates present time membrane potential $u_t$ with the previous time decayed potential $\beta u_{t-1}$.


While the MAC unit is computing the incoming potential for the current layer, the potential adder holds the accumulated potential from the previous layer. The membrane potential arithmetic unit (MAU) computes the decayed potential, and then loads the result in the potential adder.
As decayed potentials are loaded to the potential adder, the result is used to calculate the spike responses which are saved and re-read at a slower rate for MAC units, as they need the previous layer's spikes. Upon completion of reading decayed potentials, the new membrane potential values for each neuron are written as a neuron selector addresses into necessary neurons. The reset status is determined by the spike response from the Schmitt trigger module. This response determines whether a spike has occured, which determines whether the membrane potential has to be reset prior to saving or not. 


Each MAC unit can accumulate up to four incoming connections at once. This process is repeated for all fan-in spikes. 

\subsection{Control Unit}
The control unit is a state machine that has three main states: 

\subsubsection{Input Layer State}

The input layer state consumes only 3 clock cycles, as the input layer neurons each have only 1 input. It only takes one accumulation cycle for MAC units to calculate the accumulated potential, as the input spike cache is already loaded in the output layer state.
Saving results from the output layer also takes one cycle since it only has 10 output neurons and the accelerator core is designed to save up to 16 neurons in each cycle.
Initialization of the state in the control unit takes one cycle to complete.

\subsubsection{Hidden Layer State}


In this state, the MAC units iterate through 1,024 input connections for each neuron 4 connections at a time, such that 256 cycles are required to compute the potential values. While MAC units are processing, the neuron selector activates the potential adders 16 at a time, and it therefore takes 64 cycles to finish processing and storing the hidden layer potentials.

\subsubsection{Output Layer State}
In this state, only 10 MAC units are needed to compute over 256 cycles. At the same time, the input spike cache is loaded from the input spike SRAM in 8 cycles. Additionally, the potential adders will complete generating results from the hidden layer. This completes the operations needed for a single time step.

\subsection{Spike Processor}
\begin{figure}[!t]
		\centering
    	\includegraphics[width=\linewidth]{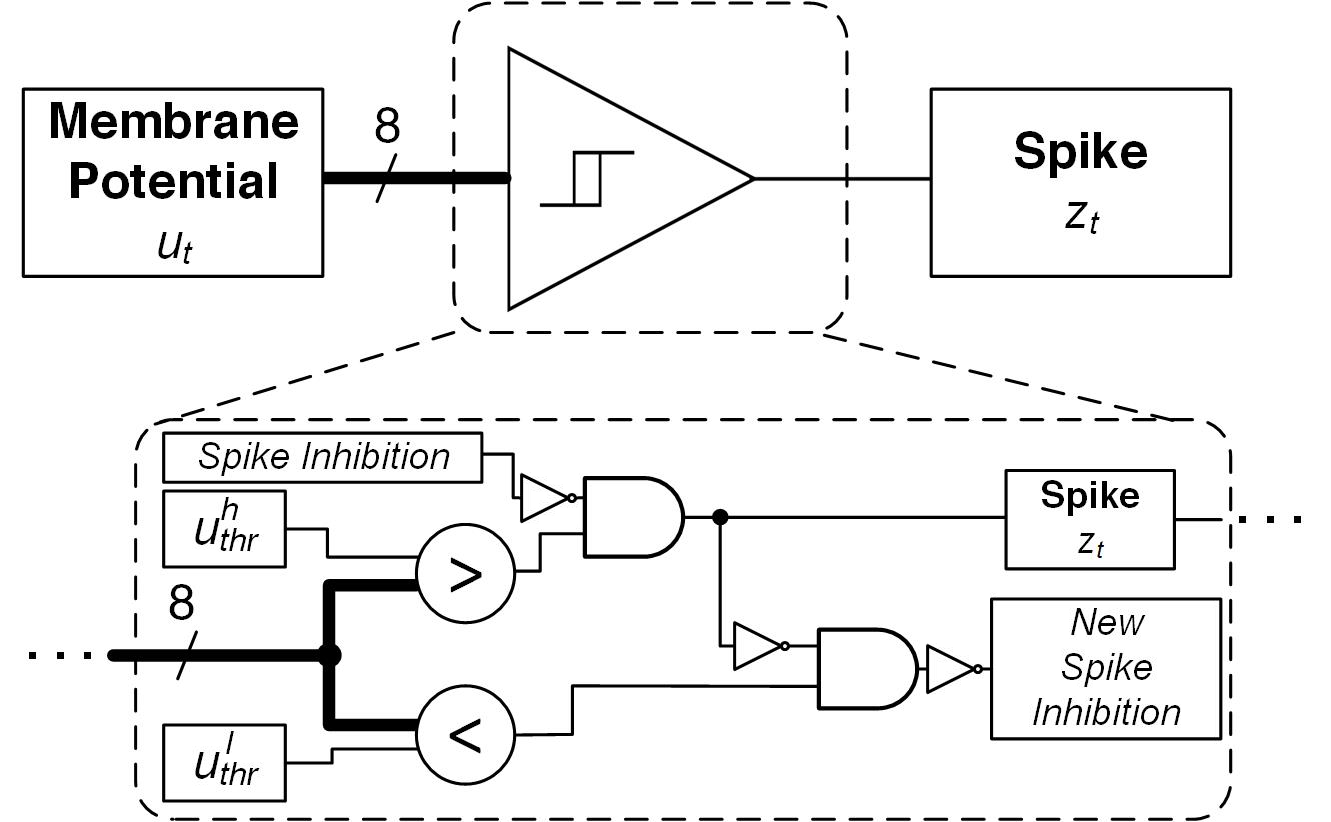}
		\caption{Spike emission process.}
		\label{fig:spike}
\end{figure}

The spike processor has two functions: i) spike emission using the Schmitt trigger as a neuron selector reads the updated membrane potential (\Cref{fig:spike}), and ii) to act as a 128-byte cache for MAC units to be used in the input layer and to be pre-loaded in the output layer state.


\subsection{Membrane Potential Arithmetic Unit}
\begin{figure}[!t]
		\centering
		\includegraphics[width=\linewidth]{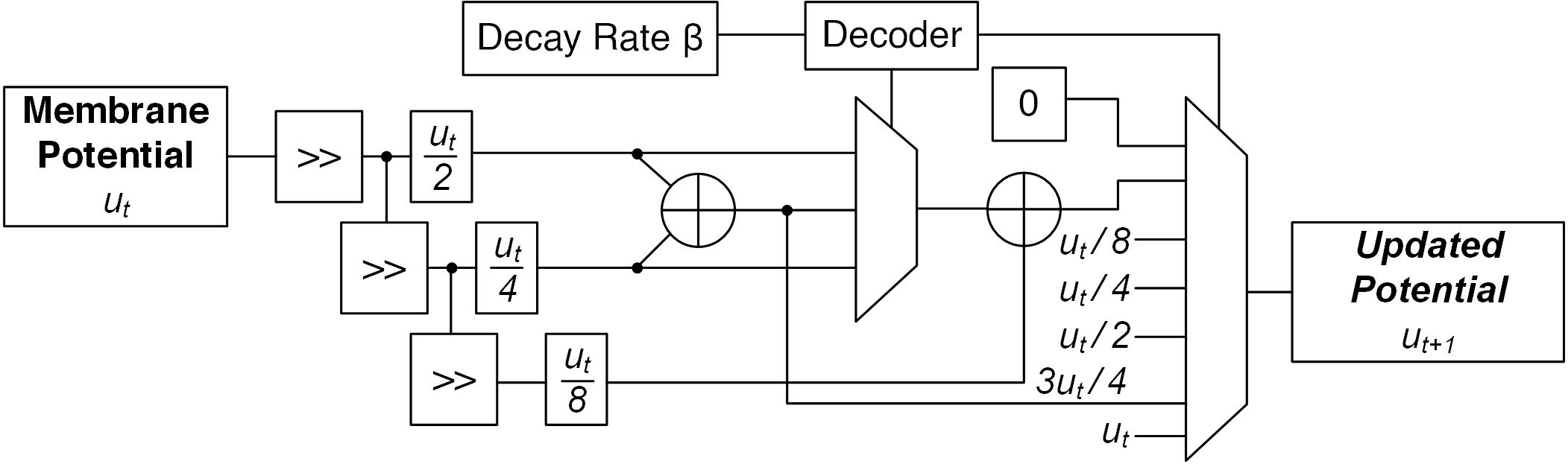}
		\caption{Membrane potential arithmetic unit uses shift-and-add to implement varying decay rates with low computational cost.}
		\label{fig:product}
\end{figure}

The MAU receives the final potential value from the neuron selector and resets the potential if that neuron has triggered a spike in the present time step. The final value is stored, and the decayed potentials  are loaded into the potential adder, where the membrane potential and the decay rate $\beta$ are read from SRAM and multiplied with a combinational circuit and routed to its respective potential adder using the neuron selector. To decrease latency, a combinational circuit was used for multiplication where the membrane potential is divided into smaller fractions using shift operations, and then they are added back together in different combinations based on the decay rate to compute the final decayed membrane potential value (\Cref{fig:product}).

\subsection{OpenRAM Macros}

Memory macros have classically been highly proprietary and manually optimized circuits, as memory cells rely on layouts that are so dense they must often bypass design rule violations. Designs are often limited by the availability of memory designs and arrays, and memory compilers are scarcely available with existing PDKs. 
OpenRAM is an open-source memory compiler that helps with addressing these issues \cite{guthaus2016openram}. 
By using a high-level Python configuration file, the layout and netlists of SRAMs can be generated by passing in data word size, the number of words in memory, for a given PDK. Process, voltage and temperature corners can also be specified for SRAM characterization.

The proposed design uses approximately 154~kB worth of dual-port 2~kB SRAM macros that consume a total area of 21.89~mm$^2$ for storing spikes, adaptive thresholds from the Schmitt triggers which enables inhibition, membrane potential, the decay rate, weights, and incoming spikes.
The SRAMs operate at 40~MHz while the core is clocked at 20~MHz. The memory controller addresses spikes to all fan-out weights, and interfaces all SRAMs to the core. The weight read lines are multiplexed within one clock cycle of the accelerator core to increase bandwidth. 


\section{Results}
\subsection{Backend Flow}
\begin{figure}[!t]
		\centering
        \includegraphics[width=200pt]{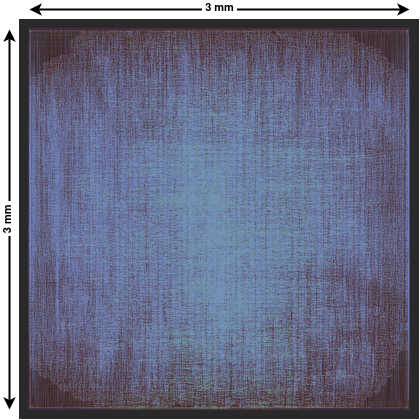}
		\caption{GDSII of the accelerator core.}
		\label{fig:gds}
\end{figure}

The OpenLane flow was used to harden the accelerator core, turning synthesized Verilog into a GDSII layout \cite{ajayi2019openroad}. Logic synthesis, floorplanning, placement, clock routing and optimization, global and detailed routing are performed within this flow. Hardening the accelerator core took 10 hours with a peak of 42.5~GB of RAM utilized, generated 128,776 physical cells, and the resulting GDSII layout is illustrated in \Cref{fig:gds}.

\subsection{Timing}
The critical path of the core is 24.56~ns due to the neuron state calculation in the MAU. To optimize the MAU, the throughput of the membrane potential decay step is doubled by including a pair of adders. The maximum clock frequency is 24.39~MHz. The SRAMs can operate at 40~MHz and take two cycles per memory access, and so to balance weight access with computation, the core is driven at 20~MHz.

\begin{table}[!t]
  \caption{Network Timing for a Dense SNN}
  \centering 
  \begin{threeparttable}
    \begin{tabular}{cc} \label{tab:late}
    Process  & Latency \\
     \midrule\midrule
    Input Layer & 0.120~$\mu$s                 \\
    Hidden Layer & 10.3~$\mu$s                \\
    Output Layer & 10.3~$\mu$s \\
     
    \cmidrule(l r ){1-2}
      Total & 20.72~$\mu$s
           \\
    \midrule\midrule
    \end{tabular}
    \begin{tablenotes}
\item[*] Network architecture: 1024-1024-10 Dense SNN.
\end{tablenotes}
\end{threeparttable}
  \end{table}

\begin{table}[!t]
  \caption{Power consumption at the fastest corner}
  \centering 
  \begin{threeparttable}
    \begin{tabular}{
    m{5em}m{3.5em}m{3.5em}m{3.5em}m{3.5em}m{2em}
    }\label{tab:power}
    Group  & Internal & Switching & Leakage & Total & \%\\
     \midrule\midrule
Sequential & 31.4~$m$W & 2.32~$m$W & 0.37~$\mu$W & 33.9~$m$W     & 28.5\%                  \\
    Combinational & 44.1~$m$W & 40.7~$m$W & 2.37~$\mu$W & 84.8~$m$W     & 71.5\%                  \\
     
    \cmidrule(l r ){1-6}
      Total & 75.5~$m$W & 43.2~$m$W & 2.74~$\mu$W & 119~$m$W    & 100.0\%                  \\
  \% &63.6\% & 36.4\% & $\sim$0.0\% &  100\%&                
           \\
    \midrule\midrule
    \end{tabular}
\end{threeparttable}
  \end{table}

\begin{table}[!t]
    \caption{Accuracy of OpenSpike across different datasets compared to full precision counterparts}
  \centering 
  
  \begin{threeparttable}
    \begin{tabular}{ccc} \label{tab:accuracy}
    & Accuracy & Accuracy
    \\
     Dataset  & (OpenSpike) & (Full precision) \\
     \midrule\midrule
    MNIST & 99.12\% & 99.42\%                 \\
    FashionMNIST & 88.12\% & 91.02\%                \\
    DVSGesture & 92.36\% & 93.06\% \\

    \end{tabular}
\end{threeparttable}
  \end{table}

The timing of various stages of the network are provided in \Cref{tab:late}, noting that this is only for one case out of any number of possible architectural configurations. Initialization involves loading firmware onto the PicoRV32 processor to configure on-chip logic analyzers, 38 general purpose IOs (maximum bandwidth of 50~MHz), and a wishbone bus to interface the accelerator with the firmware. The input layer takes 0.12~$\mu$s to complete, where input images pre-loaded on a 16kB~bank of SRAM. The input is treated as a flattened single-channel 32$\times$32 image. The input is then passed to the hidden layer, where each input is weighted by 1,024 weights and takes a total of 10.3~$\mu$s. The final layer has almost the same latency, as the same number of neuron inputs must be processed. The wallclock time for a forward-pass is 20.72~$\mu$s. The total throughput without pipelining is therefore 48,262 images per second.










\subsection{Power}
The OpenSpike core and its SRAM macros consume about the same amount of energy. SRAM macros during read and write operations of the core consume 106.54~$m$W of power in total. The dominant portion of power consumption of the core arises from dynamic power in combinational logic. A power breakdown at the fastest corner is provided in \Cref{tab:power}. Internal power refers to dynamic power in standard cells, and switching power accounts for dynamic power across routing and capacitances external to cells. The dynamic power is the total of the internal and external switching power, and amounts to 119~$m$W for a fully activated network (i.e., all inputs are spiking) in the worst-case. This can be significantly reduced by training sparse networks, e.g., by using sparse input data from event cameras \cite{amir2017low}, and by imposing objectives that aim to reduce spike count \cite{neftci2019surrogate, zenke2021remarkable, eshraghian2021training, henkes2022spiking}.

\subsection{Accuracy}
The accuracy of SNNs with binarized weights, where the network weights $w \in \{-1, +1\}$, has been measured on several image datasets, including MNIST, FashionMNIST, and DVSGesture \cite{amir2017low}. We tested the performance of small convolutional SNN architectures (16Conv5-64Conv5-10) on classification tasks in a supervised learning setup where the best performance for the MNIST dataset was 99.12\%, the FashionMNIST dataset was 88.12\%, and for DVSGesture 92.36\% was achieved (16Conv5-32Conv5-11). These represent very small performance hits when compared to their full precision counterpart networks, where MNIST only degraded by 0.3\%, FashionMNIST by 2.9\%, and DVSGesture by 0.7\%. The same hyperparameters and surrogate gradients as in Ref.~\cite{eshraghian2022fine} are adopted, where the binarization operator is replaced with a straight-through-estimator during the backward-pass.

\section{Discussion and Conclusion}
The proposed SNN accelerator core based on OpenRAM memories aims to drive neuromorphic hardware research in the direction of reproducibility, in much the same way algorithms and software development has gone. The work gone into making neuromorphic accelerators available for broad, public use are extremely useful for applications engineers in the pursuit of applied neuromorphic research, though is of less use in the chip design flow outside of algorithms exploration. Frenkel \textit{et al.} moved towards open-sourcing a neuromorphic chip for online learning, and though it relies on closed-source PDKs and memory macros, making all other aspects of the design available can help the neuromorphic community with both education and expanding the reach of custom hardware \cite{frenkel2022reckon}. Our accelerator addresses the challenges of open hardware by using a \textit{fully} open-source design flow. While open flows are presently constrained to legacy  nodes, industry involvement and more advanced technologies have already become better integrated within open-EDA toolchains within the past year, including GlobalFoundries 180~nm and SkyWater's RRAM-CMOS process. Open neuromorphic accelerators have the potential to do what the open-source community has achieved for DL. 

\bibliography{references}
\bibliographystyle{IEEEtran}

\end{document}